# 고용허가제에서 이주노동자의 최적선택 메커니즘:
## 기술적실성-자기선택 이민모형의 구축과 실증분석

이권형* · 임예진** · 조성현***

## Migrant Laborer's Optimization Mechanism Under Employment Permit System(EPS): Introducing and Analyzing 'Skill-Relevance-Self Selection' Model

Kwonhyung Lee · Yejin Lim · Sunghyun Cho

Migrant laborers subject to ROK's Employment Permit System(EPS) must strike a balance between host country's high wage and 'Depreciation of skill-relevance entailed by immigration', whilst taking account of the 'migration costs'. This study modelizes the optimization mechanism of migrant workers and the firms hiring them— then induces the solution of the very model, namely, 'Subgame Perfect Nash Equilibrium(SPNE)', by utilizing game theory's 'backward induction' method. Analyzing the dynamics between variables at SPNE state, the attained stylized facts are what as follows; [1]Host nation's skill-relevance and wage differential have positive correlation. [2]Emigrating nation's skill-relevance and wage differential have negative correlation. Both stylized facts —[1], [2]— are operationalized into 'Host nation skill-relevance hypothesis(H1)' and 'Emigrating nation skill-relevance hypothesis(H2)', respectively; of which are thoroughly tested by OLS linear regression analysis. In all sex/gender parameters(Total/Men/Women), test results support both hypotheses with statistical significance, thereby inductively substantiating the constructed model. This paper contributes to existing labor immigration literature in three following aspects: (1)Stimulate the economic approach to migrant labor analysis, and by such means, break away from the overflow of sociology, anthropology, political science, and jurisprudence in prior studies; (2)Shed a light on the EPS's microeconomic interaction process, of which was left undisclosed as a 'black box'; (3)Seek a complementary synthesis of two grand strands of research methodology — that is, deductive modeling and inductive statistics.

Keywords: Employment Permit System, self-selection of immigration, Country-specific skill-relevance

* 제1저자, 연세대학교 사회과학대학 행정학과(sangsankwon@yonsei.ac.kr)
** 제2저자, 연세대학교 언더우드국제대학 국제학과(yejinlim@yonsei.ac.kr)
*** 교신저자, 고려대학교 공과대학 산업경영공학부(cshhello99@korea.ac.kr)

# Ⅰ. 서론

2022년 현재, 저출산·고령화에 의해 가시화된 생산가능인구 감소의 실황에서, 한국의 노동이민 수용 확대는 더 이상 선택이 아닌 필수가 되었다. 그러나 이주노동에 관해 부족한, 혹은 잘못된 이해는 정책목표를 충분히 달성치 못하게 하거나, 심지어는 본래 의도와는 정반대의 효과를 야기할 수 있다. 건국 이래 그 어느 때보다도 '이주노동정책'에 대한 다각도의, 상세한 분석이 요구되는 이유이다.

이주노동정책은 크게 노동공급자(이주노동자)-중심과 노동수요자(고용주)-중심의 제도로 구분된다. 고용허가제도(Employment Permit System, EPS)는 노동수요자-중심 이주노동정책의 대표적인 예로서, '교체순환원칙'에 입각하여 근로의사가 있는 외국인력과, 인력 부족에 직면한 국내 기업을 매칭시켜주는 기능을 한다. 한국의 고용허가제는 1991년부터 시범적으로 시행된 산업연수생제도를 대체·개선하는 것으로, 내국인만으로 노동수요를 충당하지 못한 중소기업이 고용허가서를 발급받아, 고용노동부의 주재 하에, 주로 '저숙련' 외국인력을 고용할 수 있도록 하는 제도이다.[1]

우리나라의 고용허가제는 다음과 같은 과정으로 진행·운영된다. 먼저 (1)국내 기업이 임금과 근무시간, 근로조건 등을 지정한 근로계약서를 작성한 후, (2)외국인력이 매개기관을 통해 근로계약서를 확인하여 수용(체결)을 결정하면, (3)이주노동자가 지정된 직무교육을 이수, 사업장에 배치되는 순서이다. 해당 이주노동자들은 중소 제조업(예: 노동집약적 경공업 및 단순기계공업), 건설업 등 주로 '2차산업'의 노동초과수요를 충당하게 된다.

이때 이민 송출국은 대개 개발도상국이나 신흥공업국 등, 산업화를 이유로 전체 산업 대비 2차산업 규모의 비중이 높아지고 있는 국가들이다. 한편 이민 수용국은 원 거주민의 노동력이 대부분 3차산업(서비스업)에 종사하여, 흔히 '3D 업종'이라 불리는 2차산업에 종사할 원 거주민의 노동공급을 확보하기 어려우므로, 이를 고용허가제 등 이주노동정책으로 대체하는 국가이다. 요컨대, 2차산업에 특수한 직무기술은 송출국에서는 중시되나, 수용국에서는 상대적으로 경시된다고 할 것이다.

상기 (2)근로계약 수용 여부 결정의 상황에 놓인 이주노동자는, 자신의 효용을 증대(+)시키는 수용국의 (모국보다 높은) 임금과, 효용을 감소(-)시키는 '이민에 따른 보유 기술 적실성의 감가상각' 및 '이주 비용' 사이에서 조화로운 선택을 내려야 한다고 볼 수 있다. 마찬가지로, 이주노동자를 고용하는 기업 또한 이주노동자를 확충하기 위해 (즉, 근로자가 고용계약을 수용하도록 만들기 위해) 어느 수준의 임금을 어떤 방식으로 지급할지 결정을 내려야 한다.

본 연구는 고용허가제와 이주노동자의 최적선택에 관해 두 가지 상보적인 접근을 취한다. 먼저 연역적으로, 이주노동자와 기업의 최적선택 메커니즘을 '계약이론'의 방

---

[1] [고용노동부] https://www.moel.go.kr/policy/policyinfo/foreigner/list1.do (2022.08.16.).

법론을 통해 모델링하고, 경제학적 제약조건을 반영하여 모형의 최적화 문제의 해를 도출한다. 다음으로, 도출된 해답에 기반하여 가설을 수립하고, 실제 통계치의 회귀분석을 통해 수립한 가설을 검증, 모형을 귀납적으로 입증할 것이다. 이어지는 대목에서는 작금까지의 연구동향과 구축할 모형의 경제학적 배경에 관해 논한다.

## II. 선행연구 검토 및 이론적 배경

1. 이주노동 연구동향

    1) 고용허가제도에 관한 선행연구

    <표 1> 고용허가제에 관한 선행연구 분류(자모 오름차순 정리)

| 접근 | 분야 | 연구자 |
|---|---|---|
| 사회학 중심 접근 | 사회학(양적연구) | 강정향·전용일(2020), 박형기·김석호·이정환(2014), 정재도·엄광용·이재묵(2019), 하예랑(2017) 등 |
| | 사회학(질적연구)·문화인류학 | 김수경(2019), 양순미·유일상·양예숙(2018), 최경식(2019), 최서리·이창원(2021) 등 |
| | 사회정책 | 유승희(2021), 이현아(2022), 임안나(2009), 정명주·김소윤(2020), 한준성(2020) 등 |
| 정치학·행정학·정책학 중심 접근 | 정치학·행정과정 | 김영종(2009), 유정호·김민길·조민효(2017), 이기호·이화용(2015), 한준성(2022), Syifa(2017) 등 |
| | 정책분석 | 고정호(2015), 안광현·최기종(2011), 오치돈·박찬식(2010), 우영옥(2022), 유길상(2007), 윤자호(2021), 전용일·백희정(2019), Dyrberg(2015), Wanasekara(2011) 등 |
| | 비교정책 | 김용찬(2018), 배유일·주유민(2021), 전윤구(2014) 등 |
| 법학 중심 접근 | 법정책학 | 고준기(2005), 김남진(2016), 유민이(2019), 최윤철(2018)등 |
| | 인권법학 | 김성률·이원식(2017), 김종세(2021), 김종세(2015), 손윤석(2013) 등 |
| | 노동법학 | 김용환(2010), 유길상·이정혜·이규용(2004), 하갑래(2011), 최홍엽(2018) 등 |
| 경제학·경영학 중심 접근 | 노동경제학 | 김정호(2018), 문성만(2021), 서정대·박성준(2002), 정성진·김희삼(2020), 조은지·이찬영(2021), Yi&Khadka(2018)[a] |
| | 경영과학 | 김홍재·박재현·강경식(2007), 나혜숙(2008), 진현·장은미·정기선(2016) |
| 기타 | | 하재빈·이도은(2021)[b] |

a) 한국의 고용허가제도에 관한 2018년 세계은행 보고서
b) 텍스트마이닝을 통해 인터넷 상 고용허가제 키워드의 트렌드를 분석한 논문으로, <표 1>의 4가지 접근 카테고리 중 어느 하나에도 속하지 않아 '기타'로 분류하였음.

우리나라의 고용허가제(EPS)가 2004년을 전후로 도입·실시된 이후, 최근까지 다양한 학문 분과에서 각양각색의 세부 주제에 관해 광범위한 연구가 진행되었다. 이는 크게 사회학 중심/정치·행정·정책학 중심/법학 중심/경제·경영학 중심의 4가지 접근으로 분류 가능한데, 상기 <표 1>에서 확인 가능하듯이 고용허가제의 연구에 있어서 경제·경영학에 기반한 접근은 사회학, 행정학, 법학 등에 비해 상대적으로 활발히 이루어지지 않았다고 사료된다.

구체적으로, 사회학 중심의 연구에서는 고용허가제도를 통한 사회통합의 실현가능성 및 농어촌 사회에 관한 분석 등이 주를 이루었으며, 고용허가제의 사업장 변경 조항이나 사회통합기능 미비 등에 대해 비판적 논의를 공유하였다. 흥미롭게도 법학 중심 연구에서 역시 사업장 변경 조항 등 고용허가제의 인권침해적 요소에 관한 검토가

활발하였으며, 이에 따른 규범적·처방적 접근이 중심이 되었다. 한편 정치·행정학 접근에서는 산업연수생제도(1991)가 고용허가제(2004)로 변화하게 된 정치적, 제도적 경로를 추적하기도 했으며, 정책 중심 연구에서는 현황·실태진단 및 해외 이주노동 정책 사례와 비교대조 등을 통한 정책 제언이 주 축을 이루었다.

다음으로, 노동경제학적 연구에서는 고용허가제 이주노동자와 한국 내국인근로자 간 노동시장에서의 관계, 즉 노동공급 보완성-대체성의 주제가 중점적으로 분석되었다(김정호(2018), 정성진·김희삼(2020), 문성만(2021), 조은지·이찬영(2021) 참조). 특히 문성만(2021)과 조은지·이찬영(2021)은 패널 데이터를 구축하고 도구변수법을 사용하여 지역 간 노동력의 이동 등 변수를 통제, 고용허가제의 정책순효과를 측정하기도 하였다. 이외에도 서정대·박성준(2002)은 고용허가제의 실제 정책 시행에 앞서 '이주노동자에 대해 매우 비탄력적인 노동수요', '중소기업의 고용비용부담 폭증' 등 예상되는 문제점을 사전 검토하는 예방적 성격의 연구를 수행하였다.

상대적으로 노동공급자에 관한 논의가 활발한 노동경제학과 대비되어, 경영과학적 접근은 노동수요자, 즉 기업의 시각에서 고용허가제를 분석한다. 김홍재·박재현·강경식(2007)은 고용허가제를 토대로 외국인근로자에 대한 인력수급 전망 모형을 구축하여 노동수요 충당 과정의 효율화를 의욕하였으며, 나혜숙(2008)과 진현·장은미·정기선(2016)은 AMOS 소프트웨어를 사용하여 각각 고용허가제의 정책효용성(efficacy)과 당해 이주노동자의 직장 만족과 이직의사 간 상관관계를 논증하였다.[2]

### 2) 한국의 이주노동자에 관한 선행연구

본 목차에서는 상기 <표 1>의 정치학 중심 접근과 법학 중심 접근을 통합하여, 관련 문헌을 사회학 중심/정치학·법학 중심/경제·경영학 중심의 3가지로 분류하였다.[3] <표 2>에서 확인 가능하듯이, 이주노동자에 관한 선행연구에서는 고용허가제의 선행연구에 비해 각 연구접근방식 간의 불균형이 더욱더 크게 관찰된다.

구체적으로, 사회학 중심의 연구에서는 이주노동자에 대한 내국인의 태도와 인식, 이주노동자의 사회적응, 이주노동에 관한 다양한 담론(discourse)과 운동 의례, 정체성, 사회적 주변화, 타자화 등에 관해 매우 방대한 양의 양적·질적 연구가 수행되었다. 특히 사회보장·보건복지 카테고리에 속하는 문헌에서는 이주노동자의 산업안전보건에 관한 연구가 중점적으로 이루어졌으며, 더 나아가 4대보험, 복지레짐의 측면에서 배제된 이주노동자의 권리 증진 방안이 논의되기도 하였다.

---

2) Analysis of MOment Structure(AMOS); 구조방정식모형 분석에 자주 사용되는 통계 소프트웨어.
3) 즉. <표 2>에서는 정책분석 문헌이나 비교정책 문헌을 분류체계에 반영하지 않았다. 이는 '이주노동자에 관한 정책연구'가 필연적으로 대한민국의 가장 중점적인 이주노동정책인 '고용허가제'와 분리불가한 관계에 있기 때문이다. 같은 이유로 <표 1>의 법학 중심 접근과 비교하여 법정책학 문헌 역시 분류체계에 반영하지 않았다. 더해 이주노동자에 관한 정치학 기반의 연구가 시민권 등 헌법적 가치의 검토로 연장되는 경향을 보여 이를 '정치학·헌법학' 카테고리로 구체화하였다.

<표 2> 이주노동자에 관한 선행연구 분류(자모 오름차순 정리)

| | | |
|---|---|---|
| 사회학 중심 접근 | 사회학(양적연구)·인구학 | 김석호·신인철·김병수(2011), 설동훈(2009), 윤황·김해란(2011), 장호동(2022), 정현주(2020), Callinan(2020), Hasan(2011), Nguyen(2017) 등 |
| | 사회학(질적연구)·문화인류학 | 김관욱(2022), 김성숙(2011), 김연홍·임세영(2018), 김영란(2008), 백정숙·이계희(2006), 윤인진(2010), 이선화(2008), 이태정(2012), 이태정(2005), 조현미·황티·비엣하·박선주(2020), 한건수(2003), 허종호·유수영·이채정(2020), Atteraya et al.(2015), Einhorn(2022), Fan et al.(2020), Gray(2007), Skrentny et al.(2004), Ussasarn&Pradubmook-Sherer(2020), Wigglesworth&Fonseca(2016) |
| | 사회보장·보건복지[a] | 곽윤경·김기태(2021), 김기태 외 (2020), 정연·이나경(2022), 주유선(2021), 한정훈(2019), Ade(2020), Atteraya et al.(2021), Bhandari&Kim(2016), MSDP Dela Cruz(2013)[b], Korkmaz&Park(2018) 등 |
| 정치학·법학 중심 접근 | 정치학·헌법학 | 김성수(2009), 설동훈(2007), 이다혜(2014) 등 |
| | 인권·젠더법학 | 박귀천(2014), 방준식(2016), 이희성·김슬기(2015) 등 |
| | 노동법학 | 김광성(2011), 이척희·노재철(2021) 등 |
| 경제학·경영학 중심 접근 | 노동경제학·지리경제학 | 최병두(2009), 최영진(2010), 서정아(2019), Wijayasiri(2019) |
| | 경영과학 | 윤영삼·이정식(2021), 장은미·이정원(2019) |
| 기타 | | 마민(2018)[c] |

a) 상기 각주와 같은 맥락에서 '사회정책' 카테고리를 삭제하고 '사회보장·보건복지' 카테고리를 신설하였다. 사회복지학의 영역에 속하는 해당 문헌을 사회학 중심 접근에 포함하는 것에 이견이 있을 수 있으나, 다른 두 시류(strand)인 정치학·법학 중심 접근이나 경제학·경영학 중심 접근보다는 사회학 중심 접근으로 분류하는 것이 가장 적절할 것으로 사료된다.
b) 네팔의 안과의사들이 네팔 출신 한국 이주노동자들의 산업재해 특성을 분석한 논문.
c) 박범신 장편소설 『나마스테』에서 나타난 이주노동자의 적응과 시대상에 관해 분석한 문학 분야의 논문으로, <표 2>의 3가지 접근 카테고리 중 어느 하나에도 속하지 않아 '기타'로 분류하였음.

다음으로, 정치학·헌법학 중심 접근에서는 이주노동자의 정치참여 실태와 시민권, 영주권 획득 과정 등을 비판적 시각에서 검토하였으며, 노동이주에 관한 국제협력과 거버넌스 구축의 필요성을 역설하기도 하였다. 인권·젠더법학 영역에서는 미등록 외국인근로자나 체류기간 초과 이주노동자 -내지 불법체류자- 를 위한 권리구제와 함께, 여성 이주노동자, 결혼이민자 등에 관한 입법과제가 논의되었다. 노동법학 연구에서 역시 저숙련, 단기순환 이주노동에서 벗어난 '고숙련 이주노동자'의 유인 방안을 연구하는 등, 규범적·처방적 논의가 주를 이루었다.

한편 경제학 영역에서는 종전의 노동경제학 문헌과 함께, 국제 노동력 이주와 관련된 지리경제학 문헌이 특징적이었다. 보완성-대체성 논의의 연장선상에서, 최병두(2009)는 이주노동자의 유입이 지역 경제의 임금, 노동조합의 협상력, 지역생산성, 지역의 수요 등에 미치는 영향을 탐구하였으며, 서정아(2019)는 문재인 정부의 최저임금 인상 정책의제에 초점을 맞춰 '최저임금 인상'을 외생적(exogenous) 충격으로 파악하여 이주노동자와 내국인근로자의 고용감소 여부를 판단하였다. 최영진(2019)는 이주체계이론의 견지에서 동아시아 노동이주 동학(dynamic)의 모델링을 의욕하였으며, Wijayasiri(2019)는 스리랑카 출신 이주노동자의 사례를 통해 노동이주 동학의 핵심 요소인 '이주 비용'의 규모를 측정하였다.[4]

마지막으로, 경영과학적 접근에서는 주로 인력관리론에 기반한 이주노동자 연구가 진행되었다. 윤영삼·이정식(2021)은 제조업 종사 미얀마 출신 남성 이주노동자의 사례를 바탕으로 이주노동자의 우울감에 영향을 미치는 요인을 탐구하였으며, 장은미·이정원(2019)은 이주노동자의 '조직시민행동'을 위해 직장 내·작업장에서 상사와 동료의 사회적 지원이 필요함을 역설했다.

### 3) 소결

상기 연구동향 검토에서는 이주노동 문헌을 고용허가제도와 이주노동자에 관한 연구로 양분하여 살펴보았다. 주지하였다시피, 두 분야에서 모두 경제학·경영학에 기반한 접근은 사회학/정치학/행정학/정책학/법학 등 타 학문 분과에 비해 상대적으로 주목받지 못했음을 추론할 수 있다. 특히 고용허가제도 관련 정책학 문헌이나 이주노동자 연구의 사회학 문헌은 그 양의 방대함과 주제의 다양성에 있어 경제학·경영학 기반 연구를 압도하는 수준인 것으로 파악된다.

더구나 이주노동에 관한 노동경제학 기반 문헌의 경우 대다수가 이주노동자와 내국인근로자 간 '보완성-대체성 논의', 노동으로부터 자본으로의 소득재분배 현상, 정책 순효과측정 등 사후적(ex-post)이고 거시적인 차원의 연구에 한정된 경향성을 보인다. 요컨대, 고용허가제도의 경제적 '결과'가 아닌 '과정'을 해부하며, 보다 '미시적 차원'에서 이주노동자의 경제적 선택을 분석하는, 이민노동경제학 내지 이민제도경제학 연구의 새로운 접근이 요구되는 실정이라고 할 수 있을 것이다.

## 2. 이론적 배경
### 1) Roy-Borjas의 이민 자기선택 모형

일찍이 Roy(1951)는 이종산업 간 취업 여부를 결정하는 노동공급자의 자기선택(self-selection)에 관한 노동경제학적 이론의 기초를 제공하였다.[5] 이후 Borjas(1987)는 Roy(1951)의 논의를 수식으로 정형화하고, 그 함의를 이민자의 자기선택(self-selection of Immigrants)으로 확장하였는데, 논의의 핵심은 이민 여부를 결정하는 데 있어 근로자는

---

[4] 네덜란드의 이민사회학자 Hein De Hass (2010)의 이주체계이론을 의미한다. 보다 자세한 논의는 De Hass, H. (2010), "Migration and Development: A Theoretical Perspective," *International Migration Review*, Vol. 44, No. 1, p. 227-264. 참조

[5] Roy는 해당 연구를 통해 상이한 능력에 비교우위를 가지고 있는 두 집단 모두에 대한 능력 편의(ability bias)를 제거할 수 있는 방법론을 제시한 바 있다. 보다 자세한 논의는 Roy, A. D. (1951), "Some Thoughts on the Distribution of Earnings," *Oxford Economic Papers*, 3 (2): p. 135-146 참조.

$$w_1 - w_0 - M > 0 \quad (1)$$

($w_1$: 이민 수용국의 임금, $w_0$: 모국의 임금, $M$: 이주 비용(교통비, 거래 비용 등))

일 때 이민을 선택하며, 이민자의 이민 결정 시의 기대 임금이 자신의 자기 선택 메커니즘에 의존한다는 사실이다.[6] 즉 Borjas(1987)는 이주노동 '과정'을 이론화하는 데 성공한 것이다.

<그림 1> 숙련에 대한 수익 수준에 따른 양과 음의 방향의 이민 선택

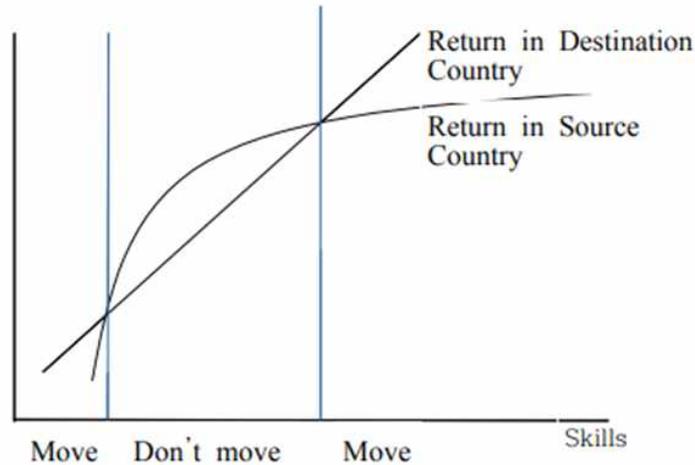

주: 음과 양 방향의 이민 선택을 돕기 위한 임금(세로축)-숙련(가로축) 평면의 숙련투자수익곡선. 원점부터 왼쪽 파란색 선까지 범위의 사람들이 '음 방향의 이민 선택'을 내리고, 오른쪽 파란색 선의 우측에 있는 범위의 사람들은 '양 방향의 이민 선택'을 내리게 된다.
자료: Borjas, G. J. and Van Ours, J. C. (2010), *Labor Economics*, p. 351, (Boston: McGraw-Hill/Irwin). 의 그래프를 자체적으로 재구성함.

구체적으로, 잠정적 이민자의 자기 선택은 두 가지로 구분할 수 있다. 양의 방향의 이민 선택(positive selection of immigrant flow)은 송출국보다 수용국의 숙련에 대한 수익 수준(returns to skill)이 높고, 숙련투자수익률 역시 더 클 때 발생한다. 뛰어난 숙련기술을 보유한 북유럽 출신의 공학자가 상대적으로 관대한(약탈적이지 않은) 소득세 수준 때문에 미국 실리콘 밸리로 이주하는 것이 이에 해당한다. 요컨대, 조세 제도의 차이로 인해 미국의 인적자본수익률이 북유럽보다 높아져, 이것이 양의 방향의 이민 선택을 야기한 것이다.

한편 음의 방향의 이민 선택(negative selection of immigrant flow)은 수용국의 숙련에 대한 수익 수준이 송출국보다 더 높긴 하나, 숙련투자수익률은 오히려 이민의 송출국이 더 클 때 발생한다. 예를 들어, 아직 사회복지의 혜택 수준이 미비한 저개

---
[6] Borjas, G. J. (1987), "Self-Selection and the Earnings of Immigrants," *American Economic Review*, 77 (4): p. 551-553. 참조

발국-개발도상국 출신의 저숙련 인력이, 미국 및 타 선진국의 (상대적으로) 관대한 공공부조 혜택에 매료되어 이민을 선택할 수 있다. 고용허가제에 따라 동남아시아 등 개발도상국에서 한국으로 저숙련 인력이 유입되는 것도 음의 방향의 이민 선택에 의한 결과로 해석 가능하다.

그런데 앞서 설시한 바와 같이, Roy-Borjas의 이민 자기선택 모형은 모형 내 두 국가에서의 소득은 근로자의 숙련 정도에만 의존한다고 가정하고 있으며, 동일한 근로자라 하더라도 이민 송출국과 수용국에서 (2차산업이 전체산업에서 차지하는 비중 등 국가의 산업구조 특성에 따라) '보유기술의 적실성'에 차이가 있다는 사실을 간과하고 있다. 다시 말해, 개발도상국에서 선진국으로 (음의 방향의) 이민을 할 때, '이민에 따른 보유기술 적실성의 감가상각'이 발생하게 되어 같은 근로자라고 하더라도 두 국가에서 동일한 숙련 수준을 가진다고 여길 수 없는 것이다.

또한 Roy-Borjas의 '일반모형'은 양과 음의 방향 이민을 모두 포함하여, 고숙련노동자와 저숙련노동자의 유입이 함께 발생하는 전통이민국가와 선발이민국가의 실정에는 적합할 수 있으나, 한국과 같이 음의 방향의 이민이 우세한 '후발이민국가'의 현실에 적용하기 어렵다.[7] 요컨대, melting-pot 국가의 '일반해'가 아니라, 신흥 이민수용국 입장의 '특수해'가 필요한 것이다. 본 연구진은 계약이론이 특수해 도출의 실마리를 제공할 수 있다고 판단하며, 이어지는 항목에서 논술한다.

### 2) 계약이론: 노동경제학에 관한 미시적 차원의 접근

계약이론(Contract Theory)은 계약, 조직, 제도에 대한 수리적 분석틀로서, 법경제학의 각론 분야로 간주되는 한편 정보경제학의 분파로 여겨지기도 한다. 이때 '구속력 있는 약속'인 '계약'과 그 당사자의 인센티브(유인)을 분석하는 것은 보통 법경제학의 연장선상으로 분류되지만, 당해 계약에서 정보 비대칭에 기인한 역선택과 도덕적 해이를 탐구하는 것은 정보경제학의 전통적 연구 주제이다.[8] 더 나아가 '계약'을 '노동계약'으로 확장하여 고용계약 참여자의 인센티브나 조직에 속한 개인의 특성을 탐구하는 것은 노동경제학과 그 맥락을 같이 한다. 요컨대, 계약이론은 미시경제학 각론의 여러 분야가 상보적으로 결합된 복합적 총체인 것이다.

---

[7] 전통이민국가란 이민으로 형성된 국가로, 미국, 캐나다, 호주, 뉴질랜드가 이에 해당한다. 선발이민국가란 1970년대 초까지 이민자가 대거 유입된 국가로, 영국, 프랑스, 독일 등 서유럽과 북유럽 국가가 대표적 사례이다. 후발이민국가는 1980년대 이후 '송출국'에서 '수입국'으로의 '이민변천'을 경험한 국가로, 스페인 등 남유럽 국가와 일본, 한국, 대만 등 동아시아 국가가 그 전형적인 예시이다. 보다 자세한 논의는 이권형·김가람·이산형 (2022), "이민정책의 현황과 과제: 이민법 체계 및 정책 네트워크의 파편화·분절화 현상을 중심으로," p. 2-3, (SK사회적가치연구원)[간행예정]. 및 이혜경 (2010), 한국 이민정책사, p. 2-4, 배재대학교, (이민정책연구원). 참조

[8] 정보경제학에서는 '행동'에 관한 비대칭 정보 문제를 도덕적 해이(moral hazard)로, '특성'에 관한 비대칭 정보 문제를 역선택(adverse selection)으로 분류한다.

계약이론에 기반한 모델링의 필수 구성요소는 '개별 합리성(Individual Rationality)'과 '유인 양립성(Incentive Compatibility)'이다.[9] 먼저 '개별 합리성'이란 계약의 (잠정적) 일방 당사자가 당해 계약에 '최소한으로 참여하기 위한' 필요조건을 의미한다. 다시 말해서, 그 계약에 참여함으로써 개인의 효용은 해당 계약에 차라리 참여하지 않는 것을 선택했을 때의 효용 수준(유보효용, Reservation Utility)보다 크거나, 최소한 그와 같은 수준이어야 한다는 것이다.

다음으로, '유인 양립성'이란, 계약이 달성하고자 하는 목표는 참여자들의 유인과 '양립 가능(compatible)'하지 않으면 안 됨을 뜻한다.[10] 대부분의 경우 계약의 쌍방, 혹은 다방 당사자들은 계약의 궁극적 목표와는 상관 없이 주어진 상황 하에서 자신의 효용을 극대화하려는 인센티브를 가지고 있다. 즉 계약의 설계자는 어떤 계약을 잠정적 참여자에게 제안하기에 앞서, 해당 당사자의 유인 체계가 어떻게 구성되어 있는지 사전에 고려하여 그것을 계약 구성에 미리 반영해야 한다는 것이다.

한편 계약이론의 모형을 해석하는 작업, 즉 계약 참여자의 최적화 문제 해결에는 '게임이론(Game Theory)'이라는 '경제학적 언어'가 활용된다.[11] 구체적으로, 2 명 이상의 당사자와 2 기 이상의 시간에 걸친 '계약'은 일종의 '순차 게임(Sequential Game)'으로 해석되며, 그러한 순차 게임의 해답은 '부분게임 완전 균형(Subgame Perfect Nash Equilibrium, SPNE)', 그리고 그 해답을 구하는 과정은 '역진귀납법(Backward Induction)'으로 풀이된다.[12]

이때 역진귀납법이란, 순차 게임(즉, '계약')의 시간상 가장 마지막 단계(최후의 부분게임)에서 거꾸로 거슬러 올라가며 부분게임 완전 균형을 찾아내는 작업을 지칭한다.[13] 상술한 개별 합리성과 유인 양립성 모두, 역진귀납법 과정에 계약 참여자의 효용함수 극대화에 대한 '제약조건(Constraint)'으로서 반영된다. 이어지는 장에서 고용주와 잠정적 이주노동자의 '고용허가제도 순차 게임'을 모형화하고, 역진귀납법을 적용하여 부분게임 완전 균형을 도출하고 그 함의를 논증할 것이다.

---

9) 김상현 (2022), "계약 및 조직이론 교안," pp. 11-13, (연세대학교).
10) 상게서, pp. 11.
11) 본 연구에서 게임이론에 관해 상세히 설명하기에는 전체 맥락상 부적절하다고 사료된다. 게임이론과 계약이론에 관한 상세한 논의는 김영세 (2022), 게임이론: 전략과 정보의 경제학, 제10판, (박영사). 참조.
12) 부분게임 완전 균형이란, 참가자들이 모든 부분게임에서 최적 선택을 구사하는 내쉬균형(an Nash Equilibrium in which every players of the game exercise the optimal choice in every subgame) 이다. 다시 말해서, 어떤 게임의 부분게임 완전 (내쉬)균형이란, 한 게임의 모든 부분게임에서 참가자들로 하여금 내쉬균형을 유도하는 전략집합(set of strategies)을 의미한다. 부분게임 완전 균형에 관한 보다 자세한 논의는 상게서, pp. 162-177. 참조.
13) 김상현 (2022), 전게서, pp. 7

## III. 모형의 구성

고용허가제를 실시하는 기업의 고용주(P, Principal의 약자)와 잠정적 이주노동자(A, Agent의 약자), 두 명의 게임 참가자(player)만 존재하는 순차게임을 생각하자.[14] P와 A는 모두 위험중립적(risk-neutral)이며, P는 A에게 성과급(piece-rate wage)만을 지급한다고 가정한다(즉, 기본급·고정급 = 0). 고용허가제를 실시하는 기업은 인력규모가 적은 중소기업이므로, P가 A의 노력($a$)을 상대적으로 매우 적은 비용으로 감시할 수 있기 때문에, 도덕적 해이의 문제는 없다고 가정한다.[15] 또한 고용노동부 및 송출국의 고용기관이 P와 A의 매개자로서 정보 전달을 용이하게 보조하므로, A의 숙련 수준에 대한 역선택 문제 역시 없다고 가정한다.

상기한 바와 같이 게임은 ①국내 기업(P)이 성과급 등을 명시한 근로 계약을 제시하면, ②잠정적 이주노동자(A)가 매개자를 통해 근로계약서를 확인하여 수용(체결)을 결정한 후, ③A가 사업장에 배치되어 최적의 노력 투입 수준을 결정하는 단계로 진행된다. 이때 ①에서 P가 제시하는 근로 계약은 최후통첩게임(ultimatum game)으로, ②에서 A는 그것을 수용할지, 거절할지의 여부만 결정할 수 있다. P와 A의 보수함수(payoff function)[내지 '효용함수']는 다음 수식 (2)-(6)과 같이 구성된다(단, 노력($a$)은 0 이상이다).

$$\text{P의 이윤: } V = R(a,\ s_1) - W_1(a) \quad (2)$$

$$\text{A가 이주할 때의 효용: } U_1 = W_1(a) - C_1(a) - M \quad (3)$$

$$\text{A가 잔류할 때의 유보효용: } U_0 = W_0(a) - C_0(a) \quad (4)$$

$$R(a,\ s_1) = s_1 a\ ,\ W_i(a) = w_i a (i = 0,\ 1) \quad (5)$$

$$C_i(a) = c_i a^2,\ c_i = \frac{1}{s_i}\ \forall i,\ \triangle w = w_1 - w_0,\ \delta = s_0 - s_1 \quad (6)$$

$$(\text{단},\ c_1 > c_0 > 0,\ 0 < s_1 < s_0,\ i \in \{0,\ 1\})$$

$W_i(a)$는 A가 $i(i \in \{0,\ 1\})$국가에서 각 국가 기업의 고용주에게 받는 임금을 의미한다. $M$은 A가 0국(송출국)에서 1국(수용국)으로 이주할 때 드는 이주 비용 혹은 교통비용, 내지 거래비용(transaction cost)에 해당한다. 앞서 설시한 바와 같이 P는 A에게 기본급/고정급 없이 성과급만을 지급하며, 각 국가에서의 성과급 임금률은 $w_i$로 나타낼 수 있다. ①단계에서 P가 근로 계약을 자신의 이윤이 극대화되도록 설계할 때, $w_1 = w_0 + \triangle w$에서 $w_0$은 국가 0의 기업이 결정하는 외생변수이므로, P의 선택 변

---

14) 고용주를 'P', 이주노동자를 'A'로 기호화한 것은 계약이론의 주인-대리인 모형(Principal-Agent Model)의 약어 체계에 의한 것이나, 본 연구에서 대리인의 도덕적 해이 등 주인-대리인 모형을 직접적으로 논하는 것은 아니다.

15) 주인-대리인 상황에서 비대칭 정보에 의한 도덕적 해이의 문제는 (1)주인이 대리인의 행동을 감시하는 비용이 높거나 (2)대리인의 부정행위(deviation)가 적발될 시 받는 처벌의 수준이 낮을 때 등의 경우 발생 가능성이 높다.

수(Choice Variable)은 △$w$, 즉 '추가 임금' 내지 '임금 차분', '임금 격차'이다.[16]

$C_i(a)$는 A가 $i(i\in\{0, 1\})$국가에서 각각 노력을 투입할 때 드는 노력 비용을 의미한다. $C_i(a)$는 $a$에 대한 이차함수로서, 볼록한 증가함수(convex increasing function)의 형태를 통해 한계비용 체증의 법칙을 반영하였다.[17] 당연하게도 $s_i$, 즉 국가-특수한 보유기술의 적실성(country-specific relevance of skill)이 증가할수록 $c_i$(노력 비용의 계수)는 감소하므로, $c_i$와 $s_i$는 서로 역수 관계를 가지는 것으로 설정하였다. 한편 국가 1에서 A의 직무는 '3D 업종'에 속해 생명과 신체에 위해가 될 수 있고, 근로환경에서 언어·문화장벽을 경험하게 되므로, 이주노동자(A)는 모국에서 일할 때에 비해 더 큰 노력 비용에 직면할 것이다. 이러한 추가적 비용을 반영하여 $c_1 > c_0 > 0$.

따라서 그 역수의 대소관계에 따라 $0 < s_1 < s_0$이 되는데, 이때 $\delta = s_0 - s_1$는 '이민에 따른 보유기술의 감가상각'을 나타내는 모수(parameter)라고 할 수 있다. 마지막으로 국가 1 고용주의 수입은 이주노동자의 노력 수준 $a$와 국가 1에서 근로자가 보유한 기술의 적실성($s_1$)에만 의존하므로(즉 P의 수입은 $s_0$과는 무관하므로), P의 수입(Revenue)을 $R(a, s_1) = s_1 a$로 구성하였다.

## IV. 최적화 문제의 해결

1. 부분게임 완전 균형의 도출

   1) 유인 양립성 조건과 개별 합리성 조건 명확화

순차게임의 풀이 방법인 역진귀납법을 통해 각각 행위자의 목적함수 극대화 문제를 해결한다. ②단계에서 이민을 결정한 이주노동자(A)는 ③단계에서 노력 수준($a$)을 선택 변수로 하여, 목적 함수(Objective Function) $U_1 = W_1(a) - C_1(a) - M$(기대 효용)을 극대화하려 할 것이다.

$$Max\ E(U_1) = W_1(a) - C_1(a) - M = (\triangle w + w_0)a - \frac{1}{s_1}a^2 - M \quad (7)$$

식 (7)의 일계조건(First Order Condition)과 이계조건(Second Order Condition)은 다음과 같다. 이계조건이 항상 음수이므로($\because 0 < s_1$), 미분을 통해 일계조건으로 유인양립성 제약조건을 구할 수 있다.

---

16) 이민 수용국(1국) 고용주가 이민 송출국(0국)의 성과급 임금률에 영향을 미칠 수 없으므로, 이민 송출국의 성과급 임금률은 내생성(endogeneity)을 가지지 않는 외생변수(exogenous variable)이다.
17) The Law of Increasing Marginal Cost. 생산량이 증가할수록 한계비용이 점차 증가하는 현상을 지칭하며, 한계비용(Marginal Cost)은 총비용(Total Cost)을 미분한 기울기와 같으므로 한계비용의 체증을 반영하려면 총비용 곡선이 볼록한 증가함수의 형태를 가져야 한다.

$$FOC: \frac{\partial U_1}{\partial a} = (\triangle w + w_0) - \frac{2}{s_1}a = 0, \quad SOC: \frac{\partial^2 U_1}{\partial a^2} = -\frac{2}{s_1} < 0 \quad (8)$$

$$\therefore a = \left\{\frac{s_1(\triangle w + w_0)}{2}\right\} \quad (IC) \quad (9)$$

②단계에서 이주노동자가 고용허가제에 참여하려면 $U_1$이 $U_0$보다 크거나, 최소한 두 효용이 같아야 하므로($U_1 \geq U_0$), 식 (10)과 같이 이민의 참여조건(개별 합리성 조건)을 구할 수 있다.

$$U_1 \geq U_0, \quad \therefore (\triangle w + w_0)a - \frac{1}{s_1}a^2 - M \geq w_0 a - \frac{1}{s_0}a^2 \quad (10)$$

식 (10)에서 좌변의 일부 항을 우변으로 넘기면 식 (11)과 같다.

$$(\triangle w + w_0)a \geq \left(\frac{1}{s_1} - \frac{1}{s_0}\right)a^2 + w_0 a + M \quad (IR) \quad (11)$$

2) 제약조건을 반영한 목적함수 극대화

P는 식 (9)의 유인 양립성 제약조건과 식 (11)의 참가유인 제약조건을 고려하여 성과급 임금률($w_i$)을 선택해야 한다. 따라서 식 (2)의 $V = R(a, s_1) - W_1(a)$는 식 (11)을 반영하여 식 (12)와 같이 재구성된다(식 (2)에 식 (11)을 대입).

$$Max\ E(V) = s_1 a - w_1 a = s_1 a - (\triangle w + w_0)a \leq s_1 a - \left(\frac{1}{s_1} - \frac{1}{s_0}\right)a^2 - w_0 a - M \quad (12)$$

이때 참여조건이 실효성 있는 제약조건(binding constraint)이 되려면, 제약조건이 등호로서 성립해야 하므로, 식 (12)의 부등식은 식 (13)의 등식으로 정리할 수 있다.[18]

$$Max\ E(V) = s_1 a - \left(\frac{1}{s_1} - \frac{1}{s_0}\right)a^2 - w_0 a - M \quad (13)$$

마찬가지로, 유인 양립성 제약조건 역시 등호로서 성립하는 것이므로, 식 (13)의 $a$에 식 (9)의 결과를 대입하여 식 (14)와 같이 재구성할 수 있다.

$$Max\ E(V) = s_1\left\{\frac{s_1(\triangle w + w_0)}{2}\right\} - \left(\frac{1}{s_1} - \frac{1}{s_0}\right)\left\{\frac{s_1(\triangle w + w_0)}{2}\right\}^2 - w_0\left\{\frac{s_1(\triangle w + w_0)}{2}\right\} - M \quad (14)$$

앞서 살펴본 바와 같이 P는 $\triangle w$를 선택 변수로써 $V$를 극대화하므로, 식 (14)를 $\triangle w$에 대해 미분한 식을 바탕으로 일계조건(FOC)과 이계조건(SOC)을 구할 수 있다. 이때 앞서 ($0 < s_1 < s_0$)의 가정에 의해 식 (15)[식(15.1)과 (15.2)]에 나타난 이계조건은 항상 음수의 값을 가지므로, 일계조건을 통해 최적화 문제를 풀 수 있다.

---

[18] 이는 귀류법(간접증명)으로 쉽게 증명할 수 있다. 쉽게 말해서, 고용주(P)는 자신의 이윤을 극대화하려는 합리적 경제인이므로, 상기 식 (12)의 부등식으로 표현된 자신의 이윤 수준의 최대치보다 미만(strictly less)인 이윤 수준을 선택할 유인이 전혀 없고, 항상 (등호로서 성립하는) 자신의 최대 이윤 수준을 선택한다는 것이다.

$$\frac{\partial^2 V}{\partial \triangle w^2}\left[s_1\left\{\frac{s_1(\triangle w+w_0)}{2}\right\}-\left(\frac{1}{s_1}-\frac{1}{s_0}\right)\left\{\frac{s_1(\triangle w+w_0)}{2}\right\}^2-w_0\left\{\frac{s_1(\triangle w+w_0)}{2}\right\}-M\right] \quad (15.1)$$

$$=\frac{s_1(s_1-s_0)}{2s_0}<0 \quad (15.2)$$

식 (14)의 일계조건을 통해 구할 수 있는 부분게임 완전 균형은 식 (16)과 같다.

$$s_0=\frac{s_1(\triangle w+w_0)}{2w_0-s_1+\triangle w} \quad (16)$$

또한 앞선 가정에 따라 $s_0>s_1$을 만족해야 하는데, 식 (16)의 변수들이 $(0<s_1<s_0)$을 만족하는 경우는 식 (17)밖에 없다.

$$w_0=0,\ 0<s_1<\triangle w,\quad s_0=\frac{s_1(\triangle w)}{\triangle w-s_1},\quad \triangle w=\frac{s_0 s_1}{s_0-s_1} \quad (17)$$

이때 식 (17)에서처럼 $w_0=0$이면 유보효용($U_0=w_0 a-\frac{1}{s_0}a^2$)이 음수가 될 수 있는데, 산업화 중인 개도국의 경우 임금인상을 억제하는 경우가 만연하므로, 현실에서도 충분히 발생 가능한 상황일 것이라고 사료된다.

2. 균형상태의 변수 간 관계 해석

최적화 문제의 해답인 식 (17)과 식 (17)이 함축하는 대소관계를 바탕으로 $\triangle w$, $s_0$, $s_1$, $\delta$ 사이의 관계를 식 (18)-(20)과 같이 정리할 수 있다.

$$\frac{\partial \triangle w}{\partial s_1}=\frac{s_0^2}{(s_0-s_1)^2}>0\ (\because 0<s_1<s_0) \quad (18)$$

$$\frac{\partial \triangle w}{\partial s_0}=-\frac{s_1^2}{(s_0-s_1)^2}<0\ (\because 0<s_1<s_0) \quad (19)$$

$$\frac{\partial \triangle w}{\partial \delta}=-\frac{s_1^2}{\delta^2}<0\ \left(\because \triangle w=\frac{s_1(s_1+\delta)}{\delta}=s_1+\frac{s_1^2}{\delta}\right) \quad (20)$$

앞서 $s_i$를 국가-특수한 보유기술의 적실성(country-specific relevance of skill)으로 정의한 바 있다. 이에 따라 식 (18)-(20)을 해석한다면 다음과 같은 정형화된 사실(stylized facts)을 알 수 있다. [1]국가 1에서 A가 보유한 기술의 적실성이 커지면 추가 임금(내지 임금 차분)은 증가한다. [2]국가 0에서 A가 보유한 기술의 적실성이 커지면 추가 임금은 감소한다. [3]이민에 따른 보유기술 적실성의 감가상각이 심화될수록, 추가 임금은 감소한다.

한편 $s_i$는 고용허가제도의 맥락에서 '2차산업'에 한정된 기술적실성이므로, '$i$국 전체 산업 구조상 2차산업이 차지하는 비중'을 그 대용치(proxy)로 사용할 수 있다. 따라서 위의 정형화된 사실 [1], [2]는 다음과 같이 재해석 가능하다. [1*]노동이민 수용국의 2차산업비중이 높을수록 이주노동자가 받는 제조업 등 2차산업 부문에서의 추가임금(임금차분)은 증가한다. [2*]노동이민 송출국의 2차산업비중이 높을수록 이민수용국-송출국 간 제조업 등 2차산업 부문에서의 임금격차가 감소한다.

이어지는 장에서 [1*], [2*]가 현실의 이주노동을 어느 정도로 잘 설명할 수 있는지 통계적으로 분석할 것이다.

## V. 가설수립 및 분석설계

1. 가설의 수립

상기 정형화된 사실 [1*]로부터 '가설 1(H1)'을 수립한다.

**H1 (이민수용국 기술적실성 가설)**: 각 노동이민 수용국의 2차산업비중과 송출국 단위 이주노동자의 수용국별 임금차분은 양(+)의 관계를 가질 것이다.

정형화된 사실 [2*]로부터 '가설 2(H2)'를 수립한다.

**H2 (이민송출국 기술적실성 가설)**: 각 노동이민 송출국의 2차산업비중과 수용국 단위 이주노동자의 송출국별 임금차분은 음(-)의 관계를 보일 것이다.

2. 변수설정 및 분석방법

H1과 H2를 검증하기 위해 다음과 같은 독립변수를 설정하였다.

**독립변수_1**: 이민수용국의 2차산업이 전체 GDP에서 차지하는 비중
**독립변수_2**: 이민송출국의 2차산업이 전체 GDP에서 차지하는 비중

다음으로, H1과 H2를 위해 각각 종속변수_1과 종속변수_2를 취하였다. 이때, 2차산업 전체의 평균임금을 측정하는 것에 있어 각국의 산업 및 노동 통계상 무엇이 2차산업의 범위에 속하는지 기준이 일치하지 않아, 논의의 편의성을 위해 '제조업 평균임금'으로 그 측정 대상을 한정하였다.[19]

---

19) 통상 2차산업에는 제조업, 광업, 건설업 등을 포함시키지만, 무엇이 제조업, 광업, 건설업에 속하는 산업인지의 세부 사항은 각국간 기준이 일치하지 않는 경우가 빈번하였다.

**종속변수_1**: 주요 이민수용국의 제조업 평균임금

그런데 H1에서 독립변수_1이 종속변수_1에 미치는 순효과를 측정하려면, 이민 송출국의 상이한 2차산업비중으로 인해 발생하는 임금차분 효과를 통제해야 한다. 다시 말해서, 다양한 수용국의 임금차분을 비교할 때 단 하나의 송출국을 기준으로 하여야 H1을 올바르게 검증할 수 있다는 것이다(ceteris paribus). 그렇다면 어떤 가상의 이민송출국 제조업 평균임금을 k라고 가정하였을 때, 해당 상수 k가 실수 범위 내의 어떠한 값을 가지더라도 종속변수_1의 서수적 분포(ordinal distribution)는 변하지 않는다.[20] 따라서 k의 값을 0이라고 하면, H1의 '단위 이주노동자의 수용국별 임금차분'은 종속변수_1의 '각 이민수용국의 제조업 평균임금'과 같은 값을 가진다.

**종속변수_2**: 주요 이민송출국과 한국 제조업 평균임금의 차분

마찬가지로, H2에서 독립변수가 종속변수에 미치는 순효과를 측정하려면, 이민 수용국의 상이한 2차산업비중으로 인해 발생하는 임금차분 효과를 통제해야 한다. 즉, 다양한 송출국의 임금차분을 비교할 때 단 하나의 수용국을 기준으로 하여야 H2를 올바르게 검증할 수 있다는 것이다. 따라서 종속변수_2에서는 '단 하나의 수용국'을 대한민국으로 설정하여, 한국 제조업의 평균임금에서 각 송출국의 제조업 평균임금을 제한 값을 취하였다. 그러나 종속변수_1과 종속변수_2를 액면 명목임금 그대로 측정하는 것은 통제변수_1을 고려하지 않은 채 유효하지 않은 분석을 강행하는 것이다.

**통제변수_1**: 각국의 물가 차이에 따른 구매력 불일치에 의한 명목임금과 실질임금 간의 괴리(discrepancy)

그러므로 본 연구에서는 종속변수_1과 종속변수_2의 제조업 명목 평균임금에 2017년도 구매력평가설(Purchasing Power Parity) 지수를 반영하여 물가 차이에 따른 측정상의 부정확성을 통제하려 의욕하였다.[21] 물론, 독립변수들과 종속변수들은 주로 2019년에 측정된 값으로서 2017년도의 물가로 국가 간 인플레이션 변수를 통제하는

---

[20] 즉, 종속변수_1은 각 수용국의 제조업 평균임금에서 k를 일괄적으로 제하는 것으로서 측정되므로, k 값의 변동에 따라 종속변수_1의 기수적 분포(cardinal distribution)는 변할 수 있어도, 그 서수적 분포(ordinal distribution)는 변하지 않는다는 것이다.

[21] 국제평가관계(International Parity Conditions)의 학설은 크게 '이자율평가설(Interest Rate Parity)'과 '구매력평가설(Purchasing Power Parity)'으로 양분된다. 전자는 국제수지상 자본금융계정 관점의 환율을 설명하는 반면, 후자는 경상수지 관점의 환율을 설명한다. 또한 전자는 '국가 간 자본이동에 의한 투자수익률은 동일해야 한다'는 가정에 기초해 있으나, 후자는 '교역재(Traded Goods)의 무위험 차익거래는 없음', 즉 '일물일가의 법칙'에 기반한다. 따라서 '제조업 및 2차산업', 다시말해 '교역재 산업'의 임금과 GDP에의 기여도를 분석하는 본 연구의 논의에서는 이자율평가설보다 구매력평가설을 기준으로 물가 변수를 통제하는 것이 보다 적절하다.

것에 2년 간 시차로 인한 괴리가 존재할 것이나, (COVID-19로 인해 노동이주가 왜곡되기 전의) 최신 이주노동 동향을 반영하려는 견지에서 2019년의 2차산업비중과 제조업 평균임금을 활용하는 것은 불가피한 선택이었다.

한편 제조업을 비롯한 2차산업은 거의 대부분의 국가에서 성비가 남초인 특성을 가지고 있으므로, 제조업의 전체 평균임금 통계는 '극단값(outlier)'으로서 여성의 제조업 평균임금을 포함할 수 있어(그리고 그에 기인하여 전체 통계가 왜곡될 수 있어), 성별 역시 다음의 통제변수_2로서 반영하여야 한다.[22]

**통제변수_2**: 각국 제조업 종사자의 성비와 당해 성별간 임금격차

따라서 본 연구에서는 종속변수_1과 종속변수_2를 각각 전체 성별 임금/남성 임금/여성 임금의 통계로 나누어, 총 6가지의 종속변수 데이터를 구축, 성비와 성별간 임금격차에 따른 측정상의 부정확성을 통제하였다.

연구진은 이어질 장에서 회귀분석 결과를 두 방식으로 제시할 것이다. 먼저 독립변수와 종속변수 간 산점도(scatter plot)를 그리고 추세선(trendline)을 표시하여 2가지 독립변수가 각각 3가지 종속변수(전체/남성/여성), 총 6가지 종속변수와 어떤 방향의 관계를 맺는지(즉, 양(+)의 방향인지 음(-)의 방향인지) 나타낸다. 이후 동일 변수들 간 OLS(최소자승법) 선형회귀분석을 실행하여 각 독립변수가 각 종속변수와 얼마나 '통계적으로 유의한' 관계를 맺고 있는지 검증한다.

그런데 이때, 퍼센트(%)로 표시된 독립변수의 단위에 비해 종속변수(2017 PPP $, 미국 달러)의 단위가 과도하게 큰 수치를 나타내고 있으므로, OLS 회귀분석에 한정하여 종속변수를 다운스케일링(down-scaling) 하였다. 구체적으로, 6가지의 종속변수에 각각 상용로그(log10)를 취하였으며, 이를 통해 단위의 차이에서 발생하는 회귀분석상의 불필요한 왜곡을 줄였을 뿐만 아니라, 로그임금의 변화를 임금률($w_i$)의 변화로 해석할 수 있게 하여 추후 분석의 편의성을 확보하였다.[23]

3. 분석 대상 국가의 선정

H1의 분석 대상 이민수용국으로는 다음 <표 3>의 국가들이 선정되었다. 구체적으로, 각 이민국가유형과 지역별로 대표적 이민수용국 16개국을 조사하여, 특정 이민국가유형 및 지역이 과대대표, 혹은 과소대표 되지 않도록 하였다.

---

[22] 제조업을 비롯한 2차산업에서는 대부분의 경우 여성의 임금이 남성의 임금보다 현저하게 낮은 (즉, 성별 간 임금격차가 큰) 특성을 보이기 때문이다.
[23] "로그임금의 작은 변화는 임금의 퍼센트 변화와 대략적으로 같다. 예를 들어, 두 직업 간에 평균 로그임금이 0.051만큼의 차이가 난다면 우리는 이 두 직업 사이에 약 5.1%의 임금차이가 존재한다고 해석할 수 있다. 이러한 특성이 노동경제학자가 임금의 로그값을 사용하여 임금에 대한 분석을 수행하는 이유 중 하나이다." Borjas, G. J. and Van Ours, J. C. (2010), 전게서, p. 13-16.에서 인용

<표 3> 분석 대상 이민수용국 주요 통계량

| 이민국가유형 | 지역 | 국가명 | 국가코드 | 이민인구비율(2019) | 연간 이민자수(2019) |
|---|---|---|---|---|---|
| 전통 | 북아메리카 | 미국 | USA | 15.4 | 50,661,149 |
| | | 캐나다 | CAN | 21.3 | 7,960,657 |
| | 오세아니아 | 호주 | AUS | 30 | 7,549,270 |
| 선발 | 서유럽 | 영국 | GBR | 14.1 | 9,552,110 |
| | | 프랑스 | FRA | 12.8 | 8,334,875 |
| | | 독일 | DEU | 15.7 | 13,132,146 |
| | 북유럽 | 스웨덴 | SWE | 20 | 2,005,210 |
| | | 노르웨이 | NOR | 16.1 | 867,765 |
| | | 핀란드 | FIN | 6.9 | 383,116 |
| 후발 | 남유럽 | 이탈리아 | ITA | 10.4 | 6,273,722 |
| | | 그리스 | GRC | 11.6 | 1,211,382 |
| | | 스페인 | ESP | 15.2 | 7,231,195 |
| | 동아시아 | 한국 | KOR | 2.3 | 1,163,655 |
| | | 일본 | JPN | 2 | 2,498,891 |
| | | 홍콩 | HKG | 39.6 | 2,942,254 |
| | | 싱가포르 | SGP | 37.1 | 2,155,653 |

주: 이혜경 (2010)과 이규용·김기선·정기선·최서리·최홍엽 (2015)이 제시한 기준에 따라 분류(단위:명). 국가코드는 ISO 3166-1 alpha-3을 기준으로 분류함
자료: UN Migrant Stock Total (2019)

H2의 분석 대상 이민송출국으로는 다음 <표 4>의 국가들을 선정하였는데, 이들은 대한민국과 고용허가제도(E-9) 협약(MOU)을 맺은 이민송출국 16개국 중 미얀마, 인도네시아, 키르기스스탄, 및 우즈베키스탄을 제외한 12개국에 해당한다.

<표 4> 분석 대상 이민송출국 주요 통계량

| 국가명 | 국가코드 | 지역 | MOU 체결일 | 일반고용허가제(E-9) 외국인근로자 현황 (2019) |
|---|---|---|---|---|
| 필리핀 | PHL | 동남아시아 | 2004. 04. | 4,575 |
| 몽골 | MNG | 동북아시아 | 2004. 05. | 785 |
| 스리랑카 | LKA | 남아시아 | 2004. 06. | 3,579 |
| 베트남 | VNM | 동남아시아 | 2004. 06. | 6,471 |
| 태국 | THA | 동남아시아 | 2004. 06. | 5,236 |
| 파키스탄 | PAK | 남아시아 | 2006. 06. | 507 |
| 캄보디아 | KHM | 동남아시아 | 2006. 11. | 7,773 |
| 중국 | CHN | 동북아시아 | 2007. 04. | 171 |
| 방글라데시 | BGD | 남아시아 | 2007. 06. | 1,646 |
| 네팔 | NPL | 남아시아 | 2007. 07. | 7,088 |
| 동티모르 | TLS | 동남아시아 | 2008. 05. | 561 |
| 라오스 | LAO | 동남아시아 | 2016. 9. | 167 |

주: 고용허가제 외국인근로자 자료 자체는 2021년까지 구성되어 있으나, COVID-19로 인해 E-9 외국인근로인력이 가파르게 급감하여 2020년 및 2021년 통계는 국가 간 비교에 적절하지 않다고 사료됨(단위:명). 국가코드는 ISO 3166-1 alpha-3을 기준으로 분류함. MOU 체결일 기준 정렬.
자료: 통계청(KOSIS), 고용허가제 외국인근로자(E-9) 국가별 도입현황, 자료갱신 2022. 02. 03.

미얀마의 경우, 제조업 평균임금이 지난 12년(2011-2022)간 총 4번밖에 측정되지 않았으며, 카렌 분쟁과 라카인 분쟁 등 내전 상황이 통계치를 왜곡할 위험이 있어 제외하였다.[24] 인도네시아 역시 2016년 이래 제조업 평균임금을 측정한 통계치가 부재하여 분석 대상에서 제외하였다. 키르기스스탄의 경우, 지난 12년(2011-2022) 간 LFS(Labor Force Survey)가 2020년, 단 한번 밖에 측정되지 않아 마찬가지로 통계의 신뢰성을 담보하기 어려워 제외하였다. 마지막으로 우즈베키스탄의 경우, OECD, 혹은 ILO의 통계치보다 다소 공신력이 낮은 정부 공식 통계(OE, Official Estimates)만 존재하며, 남성별·여성별 통계가 부재하고, 이마저도 2018년 이후로 업데이트되지 않아 제외하였다.

# VI. 회귀분석의 결과 해석

1. 가설 1(H1)의 회귀분석

앞서 설시한 바와 같이, 주요 이민수용국 16개국을 대상으로 독립변수_1과 종속변수_1(통합·남성·여성) 간 관계에 대해 최소자승법(OLS) 선형회귀분석을 수행하였다. 이하 <그림 2-1, 2-2, 3-1, 3-2, 4-1, 4-2>는 당해 통계치를 산점도와 추세선으로 시각화한 것으로, 좌측(<그림 2-1, 3-1, 4-1>) 그래프의 세로축에는 본래 단위(2017 PPP, 미국 달러($) 기준)의 종속변수_1을, 우측(<그림 2-2, 3-2, 4-2>) 그래프의 세로축에는 상용로그를 취하여 다운스케일링한 종속변수_1을 나타내었다. 각 그림에 나타난 바와 같이, 독립변수_1과 종속변수_1은 통합·남성·여성 단위 모두에서 양(+)의 관계를 보여 가설 1을 강화하였다.

<그림 2-1(좌), 2-2(우)> 독립변수_1과 종속변수_1(통합 단위)의 산점도와 추세선

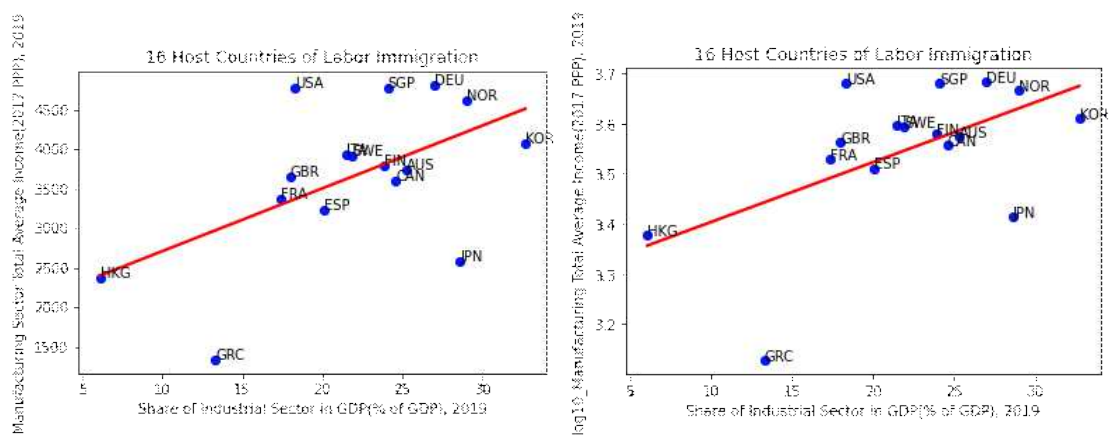

(주: 국가코드는 ISO 3166-1 alpha-3을 기준으로 분류함)

---

24) 카렌 분쟁(Karen Conflict): 1949년 1월 31일부터 카렌 주에서 벌어지고 있는 분쟁으로, 카렌 민족해방군과 미얀마군 간의 분쟁이다. 라카인 분쟁(Rakhine State Clashes): 2016년 10월 9일부터 아라칸-로힝야 구원군과 미얀마군 간 벌어지고 있는 분쟁으로, 미얀마 내부 분쟁 중 가장 최근에 발생.

<그림 3-1(좌), 3-2(우)> 독립변수_1과 종속변수_1(남성 단위)의 산점도와 추세선

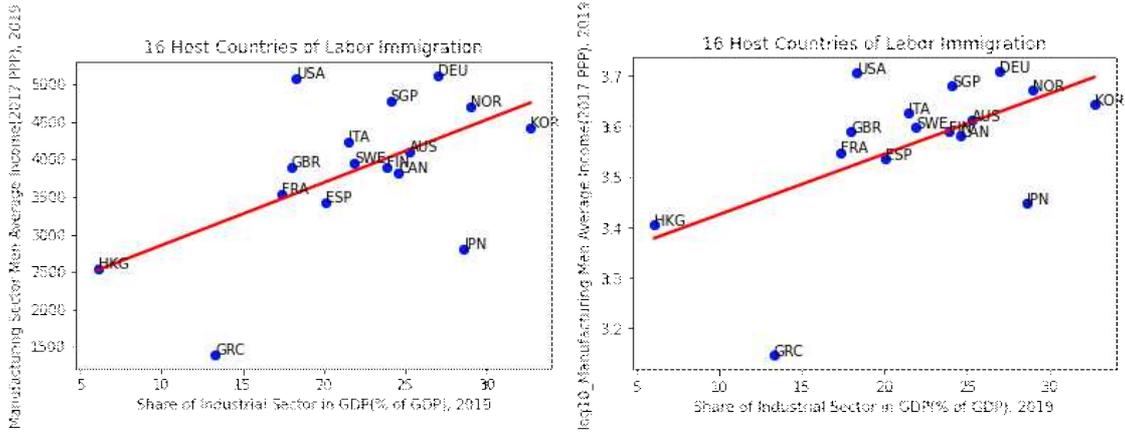

<그림 4-1(좌), 4-2(우)> 독립변수_1과 종속변수_1(여성 단위)의 산점도와 추세선

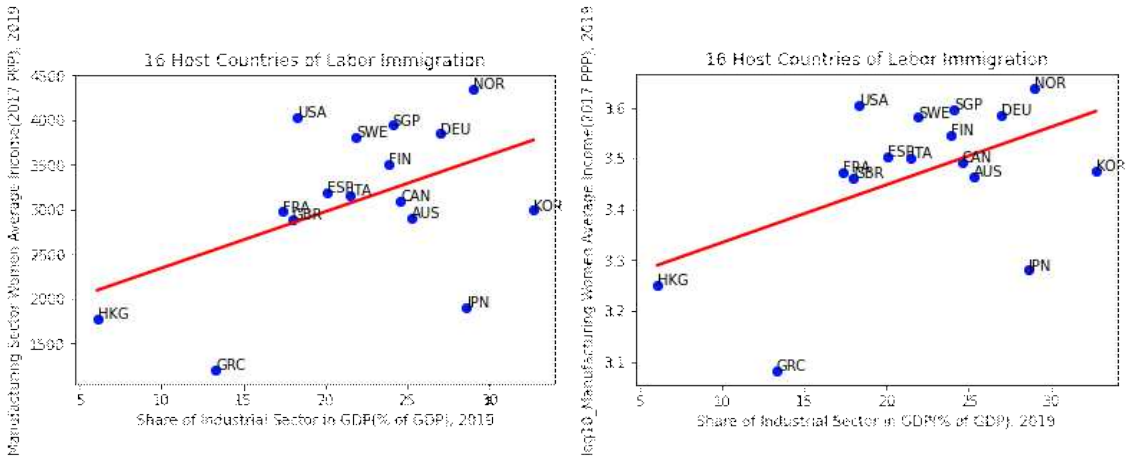

(주: 국가코드는 ISO 3166-1 alpha-3을 기준으로 분류함)

<표 5> 가설 1 OLS 선형회귀분석 결과 요약

**H1 선형회귀분석**

| Dependent Variable | R-sq | Rev. R-sq | St.Err. | F | t-value | p-value | sig |
|---|---|---|---|---|---|---|---|
| log10_수용국제조업임금 통합 | 0.303 | 0.253 | 0.122254 | 6.094 | 2.469 | 0.027 | * |
| log10_수용국제조업임금 남성 | 0.312 | 0.263 | 0.12064 | 6.362 | 2.522 | 0.024 | * |
| log10_수용국제조업임금 여성 | 0.251 | 0.198 | 0.13296 | 4.697 | 2.167 | 0.048 | * |
| Independent Variable: 2019 2차산업이 GDP에서 차지하는 비중, 세계은행 WDI(Industry) | | | | | | | |
| N = 16 | 국가: KOR, NOR, JPN, DEU, AUS, CAN, SGP, FIN, SWE, ITA, ESP, USA, GBR, FRA, GRC, HKG | | | | | | |

*p<.05, ** p<.01, *** p<.001
주: 국가코드는 ISO 3166-1 alpha-3을 기준으로 분류함, 2차산업비중 내림차순

<표 5>는 독립변수_1과 다운스케일링한 종속변수_1의 OLS 선형회귀분석 결과를 정리한 것이다. F, t-value, p-value에 나타난 바와 같이, 가설 1은 통합·남성·여성 단위 모두에서 통계적으로 유의했으며, 통합 단위의 R-square 값이 0.303으로 수용 가능한 수준의 상관성을 보였다. 또한 남성 단위의 상관계수인 0.312에 비해, 여성 단위의 상관계수는 0.251로 상대적으로 낮았다는 사실로부터 여성의 제조업 평균임금 통계에 '극단값'이 존재할 것이라는 추측이 일응 타당함을 알 수 있다.

2. 가설 2(H2)의 회귀분석

대한민국과의 고용허가제 MOU 체결 송출국 12개국을 대상으로 독립변수_2와 종속변수_2(통합·남성·여성) 사이의 관계에 대해 OLS 선형회귀분석을 수행하였다. 이하 <그림 5-1, 5-2, 6-1, 6-2, 7-1, 7-2>는 데이터를 산점도와 추세선으로 시각화한 것으로, 좌측(<그림 5-1, 6-1, 7-1>) 그래프의 세로축에는 본래 단위의 종속변수_2를, 우측(<그림 5-2, 6-2, 7-2>) 그래프의 세로축에는 상용로그를 취하여 다운스케일링한 종속변수_2를 나타내었다. 각 그림에 나타난 바와 같이, 독립변수_2와 종속변수_2는 통합·남성·여성 단위 모두에서 음(-)의 관계를 가져 가설 2를 지지하였다.

<표 6> 가설 2 OLS 선형회귀분석 결과 요약

**H2 선형회귀분석**

| Dependent Variable | R-sq | Rev. R-sq | St.Err. | F | t-value | p-value | sig |
|---|---|---|---|---|---|---|---|
| log10_임금차분통합 | 0.362 | 0.298 | 0.0319 | 5.669 | -2.381 | 0.039 | * |
| log10_임금차분남성 | 0.373 | 0.311 | 0.03185 | 5.959 | -2.441 | 0.035 | * |
| log10_임금차분여성 | 0.501 | 0.451 | 0.04227 | 10.038 | -3.168 | 0.010 | ** |
| Independent Variable: 2019 2차산업이 GDP에서 차지하는 비중, 세계은행 WDI(Industry) | | | | | | | |
| N = 12 | 국가: CHN, MNG, KHM, VNM, THA, BGD, LAO, PHL, TLS, LKA, PAK, NPL | | | | | | |

*  p<.05, **  p<.01, ***  p<.001

주: 국가코드는 ISO 3166-1 alpha-3을 기준으로 분류함, 2차산업비중 내림차순

<표 6>은 독립변수_2와 다운스케일링한 종속변수_2의 OLS 선형회귀분석 결과를 정리한 것이다. F, t-value, p-value에 나타난 바와 같이, 가설 2 역시 가설 1과 마찬가지로 통합·남성·여성 단위 모두에서 통계적으로 유의했으며, 통합 단위의 $R^2$값이 0.362로 가설 1보다도 높은 수준의 상관성을 보였다. 흥미로운 것은, 가설 1의 경우와 달리, 가설 2에서는 남성 단위의 상관계수인 0.373에 비해 여성 단위의 상관계수가 0.501로 더 높은 수치를 나타냈다는 사실이다.

<그림 5-1(좌), 5-2(우)> 독립변수_2와 종속변수_2(통합 단위)의 산점도와 추세선

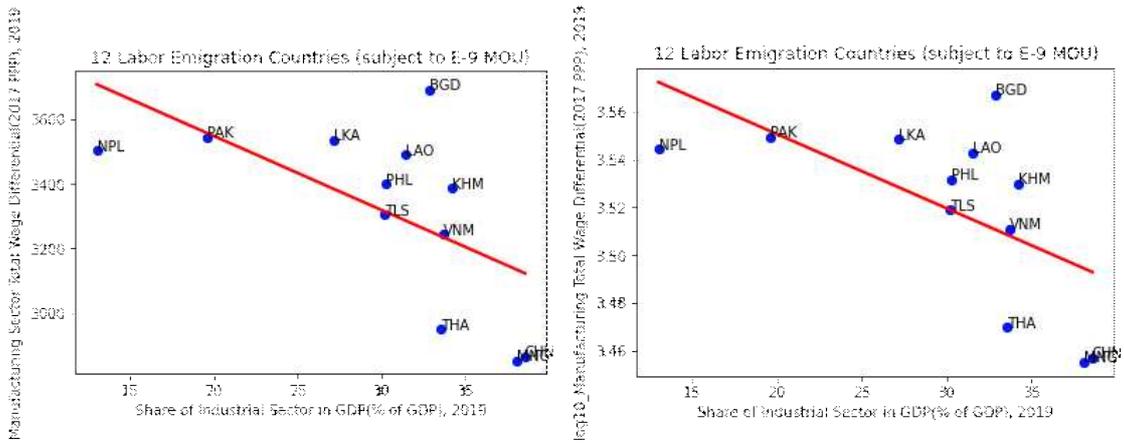

<그림 6-1(좌), 6-2(우)> 독립변수_2와 종속변수_2(남성 단위)의 산점도와 추세선

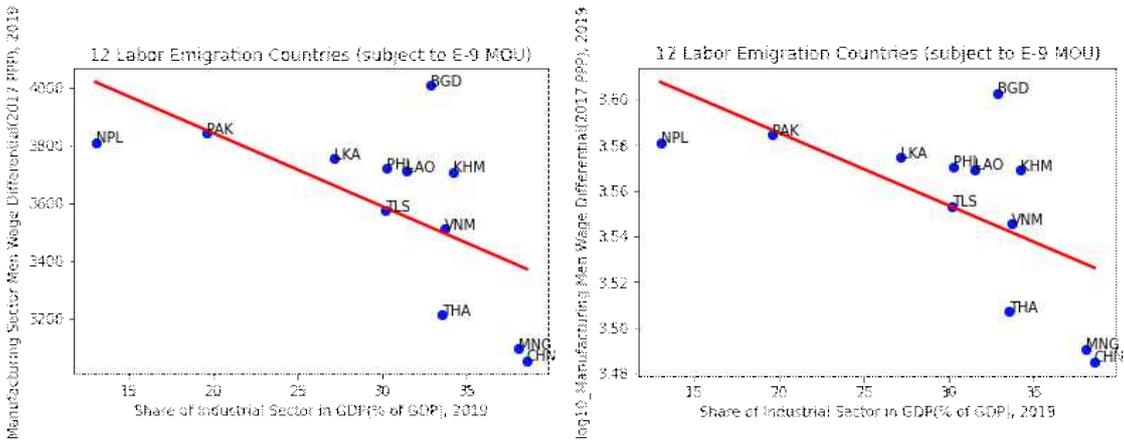

<그림 7-1(좌), 7-2(우)> 독립변수_2와 종속변수_2(여성 단위)의 산점도와 추세선

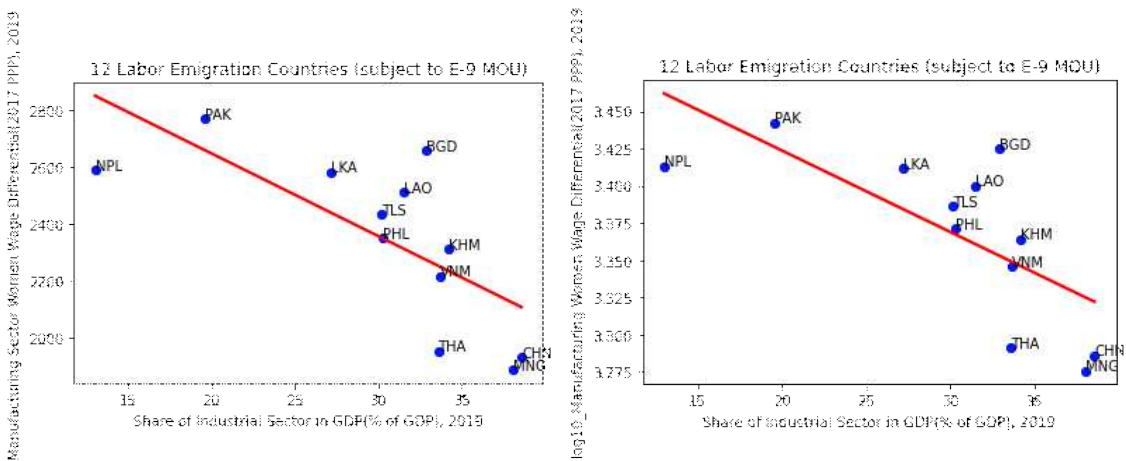

(주: 국가코드는 ISO 3166-1 alpha-3을 기준으로 분류함)

상기 두 가지의 상반된 결과는 '선진경제국(수용국)에서는 여성의 제조업 임금에 극단값이 빈번히 존재하나, 개도국(송출국)에서는 당해 극단값이 현저히 적다'로 재해석할 수 있다. 한 가지 가능한 설명은 바로 국가 내 '성별 간 임금격차'가 당해 차이를 야기한다는 것이다. 구체적으로, 개도국의 경우 성별 간 임금격차가 전체적으로 큰 수준이어서 극단값이 적은 데 비해, 선진경제국의 경우 일부 국가(특히 북유럽 국가)는 성별임금격차가 낮은 반면 다른 일부는(주로 동아시아 국가) 상대적으로 격차가 커 극단값의 빈도가 높게 표집될 수 있다. 그러나 이에 대한 자세한 원인 규명은, 아쉽지만, 논의의 전체 맥락상 추후 후속연구의 과제로 남겨 둔다.

3. 맺음말: 현실에의 재적용

지금까지 H1과 H2의 회귀분석을 통해 정형화된 사실 [1*]과 [2*]를 입증하였다. [1*]과 [2*]는 현실에 적용하였을 때 이민-노동경제학의 견지에서 상당히 흥미로운 직관(insight)을 제공한다. [1*]에 따르면, 동일한 선진경제국 국가군에 속하는 경우에도 2차산업의 비중이 더 높은 수용국일수록 잠정적 이민자가 이민에 참여할 유인이 크다는 것이다. 예를 들어, 2차산업 비중이 높은 선진국으로서 '유럽의 공장'이라 불리는 독일과, 금융 등 3차산업 비중이 높은 선진국인 영국을 비교하였을 때, 다른 조건이 같다면(ceteris paribus), 독일의 음(-)의 방향의 이민 규모가 더 클 수 있다.

또한 [2*]는 발라사-사무엘슨 효과(Balassa-Samuelson Effect)의 존재를 시사하는 정형화된 사실이다.[25] 구체적으로, 개발도상국 내지 신흥공업국에 속하는 노동이민 송출국에서 교역재 부문(2차산업)의 생산성 향상이 수용국에 비해 상대적으로 크게 발생하는 '교역재 생산혁신 충격'이 발생한 상황을 상정해 보자(단, 이때 노동이민 송출국과 수용국의 비교역재 부문(3차산업) 생산성 향상은 동일하다고 가정한다). 교역재 부문의 생산성이 향상되었으므로, 생산요소가 비교역재 부문에서 교역재 부문으로 이동할 것이고, 따라서 해당 노동이민 송출국의 전체 산업구조 대비 3차산업 비중은 감소, 2차산업의 비중은 증가할 것이다.

이때 발라사-사무엘슨 효과에 따르면, 외환시장에서 환율이 조정기능을 수행함에 따라 노동이민 송출국과 수용국의 교역재 가격은 비슷한 수준으로 수렴한다. 노동이민 송출국의 2차산업 생산성($MP_L$)이 향상되었으므로, $w = VMP_L = P^* MP_L$의 이윤극대화조건에 따라 송출국 2차산업 종사자의 임금은 증가하게 된다. 정리하자면, 교역재 부문의 생산성 향상($\triangle MP_L > 0$)이 공통원인으로서 송출국의 2차산업 비중을 높이

---

25) 발라사-사무엘슨 효과란, 교역재 부문의 생산성 향상이 발생하는 경우, 요소가 비교역재 부문에서 교역재 부문으로 이동함에 따라 비교역재 가격이 상승함으로써 실질절상이 초래되는 과정을 의미하며, 선진국-개도국 간 제조업의 임금격차는 별로 크지 않으나, 서비스업의 임금격차는 큰 현상을 설명할 수 있다는 함의가 있다. 발라사-사무엘슨 효과에 대한 보다 자세한 논의는 Mishkin, F. (2007), *The economics of money, banking, and financial markets*, p.137-145. 참조

는 동시에 수용국-송출국 간 2차산업 임금격차($\triangle w$)를 감소시키는 것이다. 이에 따라 [2*]가 설시하는 바와 같이 노동이민 송출국의 2차산업 비중 증가와 수용국-송출국 간 2차산업 임금격차($\triangle w$) 사이에는 음(-)의 상관관계가 나타나게 된다.

## VII. 결론 및 후속연구 제안

본 연구는 먼저, 계약이론에 기반하여 고용허가제 하에서 이주노동자의 최적 선택을 모델링하고, 게임이론의 방법론을 통해 부분게임 완전 균형을 도출하였다. 다음으로, 균형의 변수 간 관계가 함축하는 정형화된 사실을 가설로 구체화하고, 다각도의 회귀분석을 통해 가설을 검증하였다. 회귀분석 결과는, 모든 성별 단위에서, 통계적으로 유의미하게 가설을 지지하였으며, 이로써 앞서 구축한 모형은 귀납적으로 입증되었다.

지금까지의 논의는 이주노동 문헌에 크게 세 가지 측면에서 기여한다. 첫째, 근래까지 사회학·정치학·행정학·정책학·법학 중심의 연구에 비해 도외시되었던, 이주노동에 대한 경제학적 연구 접근을 활성화한다. 둘째, 제도의 '결과'가 아닌 '과정'에 초점을 맞춰, '암흑 상자(black box)'로서 고용허가제의 미시적 상호작용을 규명한다. 셋째, 그리고 마지막으로, 경제 이론에 입각한 연역적 모형연구와 실증 통계치 기반 귀납적 통계연구의 상호보완적 통합을 모색한다.

그러나, 유감스럽게도, 본 연구 역시 논증 과정에서 미처 다루지 못한 논의가 존재하며, 이에 관해 후속연구를 제안하며 글을 마무리한다. 먼저, 모형 구축 과정에서, 고용허가제의 '소규모 사업장' 특성에 기인하여 정보 비대칭으로 인한 도덕적 해이가 발생하지 않는다고 가정하였으나, 아무리 고용 규모가 작다고 하더라도 일정 수준의 도덕적 해이는 상존할 수 밖에 없다. 또한 고용허가제도의 매개 과정에서, 송출국의 잠정적 이주노동자가 자신의 능력(노동자 type)을 과대포장하는 등의 역선택 역시 아예 존재하지 않는다고 보기 어렵다. 종합하자면, 본 연구에서 제시한 모형에 정보 비대칭을 반영하여, 엄격하고 이상적인 가정을 완화(relax)하는 후속 연구가 요청된다고 할 것이다.

다음으로, 통계 분석 설계에서, 본 연구는 정형화된 사실 [1]과 [2]만을 가설화하였으나, 정형화된 사실 [3] 역시 마땅히 검증할 가치가 있다고 사료된다. 또한 연구진은 2019년으로 측정 시점을 고정하여, 정태적(static) 횡단면 분석만을 수행하였으나, 추후 동태적(dynamic)인 견지에서 본 모형에 관한 시계열, 혹은 패널 통계치의 분석이 이루어져야 적절하다. 마지막으로, 데이터 확보의 어려움에 기인하여, 본 연구의 종속변수는 제조업의 평균임금만으로 한정되었으나, 후속 연구에서 (비록 통일되지 않은 표준으로 인한 어려움은 감수해야겠지만) 2차산업 전체의 평균임금을 종속변수로 취한다면, 독립변수와의 관계가 더욱 명징하게 드러날 것으로 추측된다.

# 부록

<부록 1> 주요 노동이민 수용국 기술통계량(<그림 2-1 내지 4-2> 및 <표 5>)

| 노동이민<br>수용국 | 2019<br>이민인구<br>비율(%) | 2019<br>2차산업비중<br>(% GDP) | 2019 제조업<br>통합임금<br>(2017 PPP $) | 2019 제조업<br>남성임금<br>(2017 PPP $) | 2019 제조업<br>여성임금<br>(2017 PPP $) |
|---|---|---|---|---|---|
| 미국 | 15.4 | 18.3 | 4778.4 | 5081 | 4031.1 |
| 캐나다 | 21.3 | 24.6[a] | 3611.2 | 3811.8 | 3101.3 |
| 호주 | 30 | 25.3 | 3743.8[b] | 4105.1 | 2906.3 |
| 영국 | 14.1 | 18 | 3658.3 | 3889.4 | 2893 |
| 프랑스 | 12.8 | 17.4 | 3381.2[c] | 3534.7[d] | 2976.2[e] |
| 독일 | 15.7 | 27 | 4812.7 | 5112.4 | 3847.8 |
| 스웨덴 | 20 | 21.9 | 3914.8 | 3956.6 | 3810.4 |
| 노르웨이 | 16.1 | 29 | 4630.1 | 4707.6 | 4346.9 |
| 핀란드 | 6.9 | 23.9 | 3799.7 | 3897.1 | 3512.9 |
| 이탈리아 | 10.4 | 21.5 | 3942.3 | 4239.9[f] | 3164.4[g] |
| 그리스 | 11.6 | 13.3 | 1343.8 | 1399.5 | 1210.3 |
| 스페인 | 15.2 | 20.1 | 3239.6 | 3433.7 | 3194.4 |
| 한국 | 2.3 | 32.7 | 4073 | 4411.7 | 2992.1 |
| 일본 | 2 | 28.6 | 2586.3 | 2802.1 | 1908 |
| 홍콩 | 39.6 | 6.1[h] | 2383.5[i] | 2542.4[j] | 1779.7[k] |
| 싱가포르 | 37.1 | 24.1 | 4778.4 | 4786 | 3946 |

a) 2018년 통계치로 대체(2019년 조사되지 않음).
b) 상기 각주와 내용 같음.
c) 2014년 통계치로 대체(ADM - Déclaration Annuelle de Données Sociales 데이터는 2015년부터 2020년까지 비교적 연속적으로 조사되어 왔으나, 고용주 입장에서 각 직원에 대한 보수 금액을 신고하는 것이기 때문에 그 임금 수준이 실제보다 높게 집계되는 경향이 있어 제외함; 따라서 ES - Enquête Annuelle sur le coût de la main d'oeuvre et la Structure des salaires의 최신치인 2014년의 통계를 반영함, ES는 2015년 이후로 조사되지 않음).
d) 상기 각주와 내용 같음.
e) 상기 각주와 내용 같음.
f) 2013년 통계치로 대체(ES - Labour-related Establishment Survey 데이터는 2010년부터 2020년까지 비교적 연속적으로 조사되어 왔으나, 성별 단위 통계가 부재하고 전체 통합 단위의 조사만 진행됨; 따라서 성별 단위가 데이터가 존재하는 HIES - EU Statistics on Income and Living Conditions의 최신치인 2013년의 통계를 반영함, HIES는 2014년 이후 조사되지 않음).
g) 상기 각주와 내용 같음.
h) Official Estimates(홍콩 통계청), 2020년 통계로 대체(2019년 조사되지 않음).
i) 2016년 통계치로 대체(2017년 이후로 조사되지 않음).
j) 상기 각주와 내용 같음.
k) 상기 각주와 내용 같음.

주: 해당 연도의 통계치가 미비된 경우에는 2019년과 최대한 가까운 연도의 통계치 값으로 대체함.

자료: ILO Worldwide Income Database, United Nations International Migrant Stock 2019 Database, World Bank World Development Index

<부록 2> 고용허가제도 송출국 기술통계량(<그림 5-1 내지 7-2> 및 <표 6>)

| 노동이민 송출국 (E-9 도입국) | 2019 E-9 외국인근로자 현황(명) | 2019 2차산업비중 (% GDP) | 2019 제조업 통합임금 (2017 PPP $) | 2019 제조업 남성임금 (2017 PPP $) | 2019 제조업 여성임금 (2017 PPP $) |
|---|---|---|---|---|---|
| 필리핀 | 4,575 | 30.3 | 673.2 | 692.7 | 641.1 |
| 몽골 | 785 | 38.1 | 1220 | 1316.8 | 1104.7 |
| 스리랑카 | 3,579 | 27.2 | 537.5 | 658.8 | 409 |
| 베트남 | 6,471 | 33.72 | 830.4 | 902[a] | 774.2[b] |
| 태국 | 5,236 | 33.6 | 1123.1 | 1198.5 | 1037.1 |
| 파키스탄 | 507 | 19.6 | 529.4 | 571.7 | 221.4 |
| 캄보디아 | 7,773 | 34.2 | 687.2 | 705.7 | 678 |
| 중국 | 171 | 38.6 | 1207.8[c] | 1356.36[d] | 1059.24[e] |
| 방글라데시 | 1,646 | 32.9 | 382.5[f] | 406[g] | 331.8[h] |
| 네팔 | 7,088 | 13 | 571.5[i] | 601.6[j] | 401.1[k] |
| 동티모르 | 561 | 30.2 | 770.1[l] | 838[m] | 556[n] |
| 라오스 | 167 | 31.5 | 583.2[o] | 703.1[p] | 479.6[q] |
| 한국 | N/A | 32.7 | 4073 | 4411.7 | 2992.1 |

a) Official Estimates(베트남 노동보훈사회부)
b) 상기 각주와 내용 같음.
c) 2016년 통계치로 대체(2017년 이후로 조사되지 않음).
d) 2016년 통계치로 대체(2017년 이후로 조사되지 않음). 중국 정부 당국에서 제조업 부문의 남-녀 평균 임금에 관한 통계치를 일체 발표하지 않고 있음; 이에 梁勤(liángqín) & 许东黎(xǔdōnglí) (2018)의 연구결과와 Wei X., Ma, E., and Wang, P. (2017)의 연구결과에 기반하여 임금비율 및 격차 추정 후 반영하였음.
e) 상기 각주와 내용 같음.
f) 2017년 통계치로 대체(2018년 이후로 조사되지 않음).
g) 상기 각주와 내용 같음.
h) 상기 각주와 내용 같음.
i) 상기 각주와 내용 같음.
j) 상기 각주와 내용 같음.
k) 상기 각주와 내용 같음.
l) 2016년 통계치로 대체(2017년 이후로 조사되지 않음).
m) 2013년 통계치로 대체(2016년에 조사되었으나, 극단값이어서 제외).
n) 상기 각주와 내용 같음.
o) 2017년 통계치로 대체(2018년 이후로 조사되지 않음).
p) 상기 각주와 내용 같음.
q) 상기 각주와 내용 같음.

주: 해당 연도의 통계치가 미비된 경우에는 2019년과 최대한 가까운 연도의 통계치 값으로 대체함.
자료: ILO Worldwide Income Database, 통계청 고용허가제 외국인근로자(E-9) 국가별 도입현황, World Bank World Development Index

# 고용허가제에서 이주노동자의 최적선택 메커니즘:
## 기술적실성-자기선택 이민모형의 구축과 실증분석

이권형·임예진·조성현

고용허가제 하의 이주노동자는 수용국의 높은 임금과, '이민에 따른 기술적실성의 감가상각' 및 '이주 비용' 사이에서 조화로운 선택을 내려야 한다. 본 연구는 고용허가제 하 이주노동자와 기업의 최적선택 메커니즘을 '계약이론'으로 모델링하고, 게임이론의 역진귀납법을 통해 부분게임 완전 균형을 도출한다. 균형 상태의 변수 간 관계로부터 [1]수용국의 보유기술적실성과 임금 차분은 양(+)의 관계이며 [2]송출국의 보유기술적실성과 임금 차분은 음(-)의 관계임이 추론된다. [1]과 [2]로부터 '이민수용국 기술적실성 가설'과 '이민송출국 기술적실성 가설'을 수립하며, OLS 회귀분석으로 가설을 검증한다. 회귀분석 결과는 모든 성별 단위에서 통계적으로 유의미하게 가설을 지지하며, 이로써 구축한 모형을 귀납적으로 입증한다. 본 연구는 이주노동에 관한 경제학적 접근을 활성화하며, '암흑 상자'로서 고용허가제의 미시적 상호작용 과정을 규명하고, 연역모형 및 귀납통계 연구의 상보적 합일을 모색한다.

[주제어: 고용허가제, 이민의 자기선택, 기술적실성]